\documentclass[prl, twocolumn,preprintnumbers,amsmath,amssymb,showpacs,aps,pra]{revtex4}
\usepackage{mathrsfs}

\usepackage{graphicx}
\usepackage{dcolumn}
\usepackage{amsmath}
\usepackage{epsfig}
\usepackage{subfigure}
\usepackage{color}
\usepackage{epstopdf}
\usepackage{framed}

\begin{document}

\date{\today}

\title{Ultrabroadband Quantum-Secured Communication}%
\author{Quntao Zhuang}
\email{quntao@mit.edu}
\author{Zheshen Zhang}
\author{Justin Dove}
\author{Franco N. C. Wong}
\author{Jeffrey H. Shapiro}
\affiliation{Research Laboratory of Electronics, Massachusetts Institute of Technology,
77 Massachusetts Avenue, Cambridge, Massachusetts 02139, USA}%

\begin{abstract} 
We propose a two-way secure-communication protocol in which Alice uses an amplified spontaneous emission source while Bob employs binary phase-shift keying and an optical amplifier.  Against an eavesdropper who captures all the light lost in fibers linking Alice and Bob, this protocol is capable of 3.5\,Gbps quantum-secured direct communication at 50\,km range.  If Alice augments her terminal with a spontaneous parametric downconverter and both Alice and Bob add channel monitors, they can realize 2\,Gbps quantum key distribution at that range against an eavesdropper who injects her own light into Bob's terminal.  Compared with prevailing quantum key distribution methods, this protocol has the potential to significantly increase secure key rates at long distances by employing many ultrabroadband photons per key bit to mitigate channel loss.

\end{abstract}


\maketitle

Quantum key distribution (QKD) allows a random bit-string to be shared between remote parties with security vouchsafed by the laws of quantum mechanics \cite{Bennett1984,Ekert1991,Gisin2002,Grosshans2002}.  Were its secret-key rate equal to the desired communication rate, QKD would enable full information-theoretic security by means of one-time pad encryption.  Unfortunately, demonstrated secret-key rates fall far short of what is needed for widespread use of one-time pad encryption.  At present, therefore, QKD's principal application is to rekey classical cryptosystems, making overall security that of the classical system.   In this regard, quantum-secured direct communication (QSDC) \cite{Bostrom2002,Deng2004} may be preferable to key distribution, especially if it can be done at high rates.  Quantum illumination (QI) \cite{Shapiro2009a,Zhang2013,Shapiro2014} is an entanglement-based two-way protocol that has the potential to accomplish that goal, but in its present form has two major drawbacks, which will be described below.  In this Letter we report a new protocol, based on QI, that eliminates these drawbacks and offers unprecedented rate-versus-distance capability.  The new protocol realizes passive eavesdropping immunity---i.e., immunity to an Eve who merely listens to light leaking out of the connections between Alice and Bob---without the use of entanglement.  That immunity is shown to arise from the no-cloning theorem, as is the case for the security of both discrete-variable and continuous-variable QKD.  Entanglement does come into play, however, in making the new protocol resistant to active eavesdropping, i.e., when Eve injects her own light into Bob.  We begin our development with a description of the QI protocol as it provides the foundation for what will follow.

In QI, Alice creates entangled signal and idler beams and transmits the signal to Bob while retaining the idler.  Bob modulates the light he receives from Alice using binary phase-shift keying (BPSK), amplifies the modulated light to boost the signal strength and mask his message with amplified spontaneous emission (ASE) noise, and returns the resulting light to Alice.  Owing to the initial entanglement, Alice can recover Bob's message while a passive eavesdropper cannot.  The first drawback of QI is its vulnerabilty to active eavesdropping, which is a major concern for QI carried out over optical fibers.  Nevertheless, we will postpone discussing how active eavesdropping might be defeated until we address QI's other drawback, which limits its rate-versus-distance capability even when Eve only employs passive eavesdropping.

Suppose that the protocol from \cite{Shapiro2009a} is used on a fiber channel, as shown in Fig.~1.  Here: (1) Alice employs a continuous-wave (cw) spontaneous parametric downconverter (SPDC) source, with phase-matching bandwidth $W$ and brightness (average photon number/s-Hz) $N_S \ll 1$, transmits the signal beam to Bob over a fiber link of transmissivity $\kappa_S \ll 1$, and stores her idler beam in a fiber loop of transmissivity $\kappa_I$; (2) Bob uses $R\,$bps BPSK to encode his message and an erbium-doped fiber amplifier (EDFA) 
\begin{figure}
\includegraphics[width=3.25in]{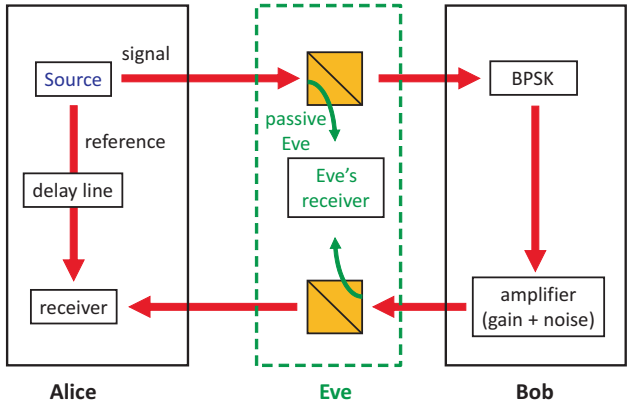}
\caption{(color online).  Setups for QI-secured direct communication in the presence of passive eavesdropping.  In the protocol from \cite{Shapiro2009a}, Alice's source is a continuous-wave spontaneous parametric downconverter whose signal she sends to Bob and whose idler she retains as a reference for the optical parametric amplifier receiver from \cite{Guha2009}.  For the new protocol, Alice employs a source of amplified spontaneous emission noise to send a low-brightness signal to Bob while retaining a high-brightness copy for use as the local oscillator in a broadband balanced-homodyne receiver.}
\end{figure} 
of gain $G_B\gg 1$ and ASE output noise of brightness $N_B = G_B$ to mask his message; and (3) Alice recovers Bob's message with the best known receiver for that task, viz., the optical parametric amplifier (OPA) receiver \cite{Guha2009}.  The OPA receiver requires Alice to store her idler for a time equal to the roundtrip propagation delay between Alice and Bob, making $\kappa_I = \kappa_S^2$.  The error probability of Alice's receiver then satisfies $\Pr(e)_{\rm Alice}^{\rm OPA} \sim \exp(-2W\kappa_S^3G_BN_S/RN_B)/2$, assuming ideal equipment \cite{Shapiro2014,footnote1}.   In contrast, Eve's passive-eavesdropping error probability for her (realization unknown) optimum quantum receiver obeys $\Pr(e)_{\rm Eve}^{\rm opt} \sim \exp(-4W\kappa_SG_BN_S^2/RN_B)/2$ \cite{Shapiro2009a,footnote1}, assuming that she collects all the light lost en route between Alice and Bob.  The $\kappa_S^3N_S$  versus $\kappa_SN_S^2$ disparity between Alice and Eve's error exponents provides Alice and Bob with direct-communication security against this passive-eavesdropping individual attack when the former substantially exceeds the latter \cite{footnote2}.  Alice's $N_S$ versus $N_S^2$ advantage derives from the maximally-entangled nature of the phase-sensitive cross correlation between the signal and idler emitted by her SPDC source, while her $\kappa_S^3$ versus $\kappa_S$ disadvantage is due to her needing an idler-storage fiber that is twice as long as the Alice-to-Bob fiber.   

An alternative QI architecture that avoids this idler-loss scaling was introduced in \cite{Shapiro2014}. It replaces Bob's single-channel BPSK modulator with $K$ independent dense wavelength-division multiplexed (DWDM) BPSK modulators that encode rate $R_c$\,bps data on nonoverlapping $W_c$\,Hz slices of the signal spectrum, and it replaces Alice's OPA receiver with $K$ DWDM homodyne receivers for the matching spectral bands of her idler and the light returned to her from Bob.  Because homodyne detection of the idler can be done immediately, with delay-locking performed post-detection, Alice's error probability for each DWDM channel is given by $\Pr(e)_{\rm Alice}^{\rm DWDM} \sim \exp(-W_c\kappa_SG_BN_S/R_cN_B)/2$ with ideal equipment \cite{Shapiro2014,footnote1}.  Unfortunately, the implementation burden for this architecture is overwhelming:  50 DWDM channels with $W_c = 40\,$GHz bandwidth would be needed to fully populate a $W=2$\,THz phase-matching bandwidth, and each of those channels would need a high-speed analog-to-digital converter to retain the information required to realize the error probability given above.  Thus, the DWDM architecture is \em not\/\rm\ a satisfactory solution to the idler-storage problem.  In its stead we propose replacing Alice's SPDC source with an ASE source (e.g., an EDFA), and her OPA receiver with a  broadband (single-channel) balanced-homodyne receiver.  The ASE source's high brightness permits
a low-brightness portion of its output to be sent to Bob; such light can perfectly mimic what Bob (and a passively eavesdropping Eve) would see were Alice to have used an SPDC source.  However, because Alice now retains high-brightness ASE light that is perfectly correlated with what she is sending to Bob, fiber-spool storage for the roundtrip delay is no longer problematic, i.e., she can use optical amplifiers in her storage system with insignificant error-exponent (signal-to-noise ratio) penalty \cite{footnote3}.  Moreover, this bright stored light is exactly what is needed as the local oscillator (LO) for broadband homodyne reception of the light returned by Bob.  Now, Alice's error probability, with ideal equipment, is given by $\Pr(e)_{\rm Alice}^{\rm ASE} \sim \exp(-W\kappa_SG_BN_S/RN_B)/2$ \cite{footnote1,footnote3}, while the passive-eavesdropping Eve's optimum quantum receiver's performance is still $\Pr(e)_{\rm Eve}^{\rm opt} \sim \exp(-4W\kappa_SG_BN_S^2/RN_B)/2$.  Comparing the asymptotic expressions for $\Pr(e)_{\rm Alice}^{\rm OPA}$ and $\Pr(e)_{\rm Alice}^{\rm ASE}$ reveals that the latter loses a factor of two in its error exponent compared to the former, but gains a much larger benefit by virtue of its immunity to the idler-storage loss that plagues the SPDC-based system.  Another benefit afforded by Alice's switching to an ASE source is that she can optimize her secure rate by choosing the brightness of the signal light she sends to Bob over a much wider range than is available from an SPDC source.  This is because the bulk-optics cw SPDC sources that produce maximally-entangled signal and idler outputs are generally limited to $N_S \le 10^{-3}$, but commercial EDFAs can yield output brightness as high as $10^4$, from which any desired $N_S \le 1$ signal brightness is easily obtained.  We also note that the upper bound from \cite{Zhang2013} on Eve's passive-eavesdropping Holevo information as a function of Alice's $N_S$ also applies, unchanged, when Alice uses an ASE source.  

Figure~2(a) compares QI's secure-rate-versus-distance performance when Alice uses the ASE-source/homodyne-receiver setup proposed herein, in lieu of the SPDC-source/OPA-receiver configuration from \cite{Shapiro2009a}, and a passive-eavesdropping Eve mounts an optimum collective attack.  The curves therein are secure-rate lower bounds for the SPDC and ASE systems obtained from 
$\Delta I_{AB}^{\rm LB} = \beta I_{AB} - \chi_{EB}^{\rm UB}$, where $I_{AB}$ is Alice and Bob's Shannon information rate, $\chi_{EB}^{\rm UB}$ is an upper bound on Eve's Holevo information rate, and $\beta$ is the efficiency of the error-correcting code employed by Alice and Bob \cite{footnote3}.  
Both curves assume that: (1)  Alice and Bob are connected by fibers with 0.2\,dB/km loss; (2) Alice's source has 2\,THz bandwidth; (3) the brightness $N_S$ of the signal light Alice sends to Bob and the bit rate $R\le 10$\,Gbps of Bob's BSPK modulator are chosen to maximize $\Delta I_{AB}^{\rm LB}$; (4) Bob's amplifier has gain $G_B = 10^4$ and output ASE with brightness $N_B = 10^4$; and (5) Eve collects all light lost en route between Alice and Bob.  For the SPDC/OPA system, Alice stores her idler in fiber with 0.2\,dB/km loss, and she chooses her OPA's gain $G_A$---in conjunction with her $N_S$ and Bob's $R$---to maximize $\Delta I_{AB}^{\rm LB}$. Her receiver's photodetector quantum efficiency is 0.9, its bandwidth is sufficient to accommodate 10\,GHz BPSK, and $\beta = 1$ is assumed.  For the ASE/homodyne system, Alice's reference---i.e., her $N_{\rm LO} = 10^4$ brightness LO---is presumed to be stored without degradation, her homodyne receiver's efficiency is 0.9 \cite{footnote4}, and its bandwidth is sufficient for 10\,GHz BPSK.  In addition, her error probability is constrained to satisfy $\Pr(e)^{\rm ASE}_{\rm Alice} \le 0.1$, making her $\beta = 0.94$ achievable at $\Pr(e)^{\rm ASE}_{\rm Alice} = 0.1$ with a rate-1/2 low-density parity-check code \cite{Richardson2001}.  All other aspects of both systems are ideal.  
\begin{figure}
\includegraphics[width=1.6in]{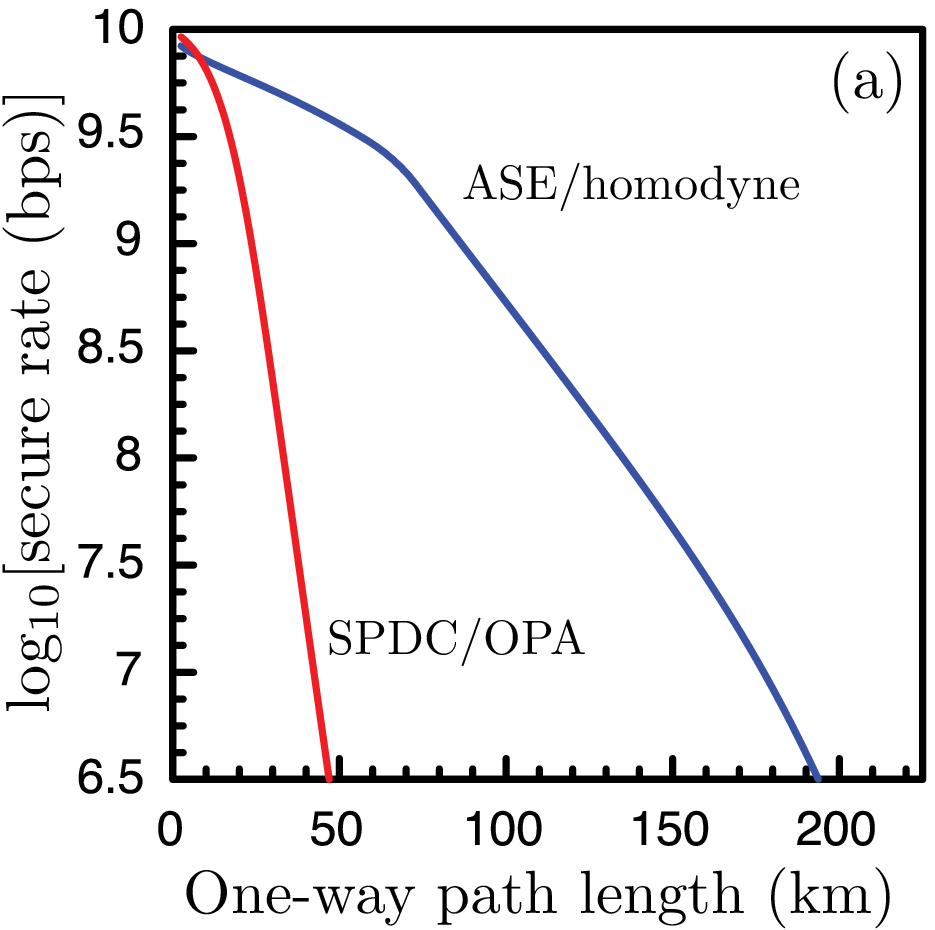}\hspace{.2in}\includegraphics[width=1.6in]{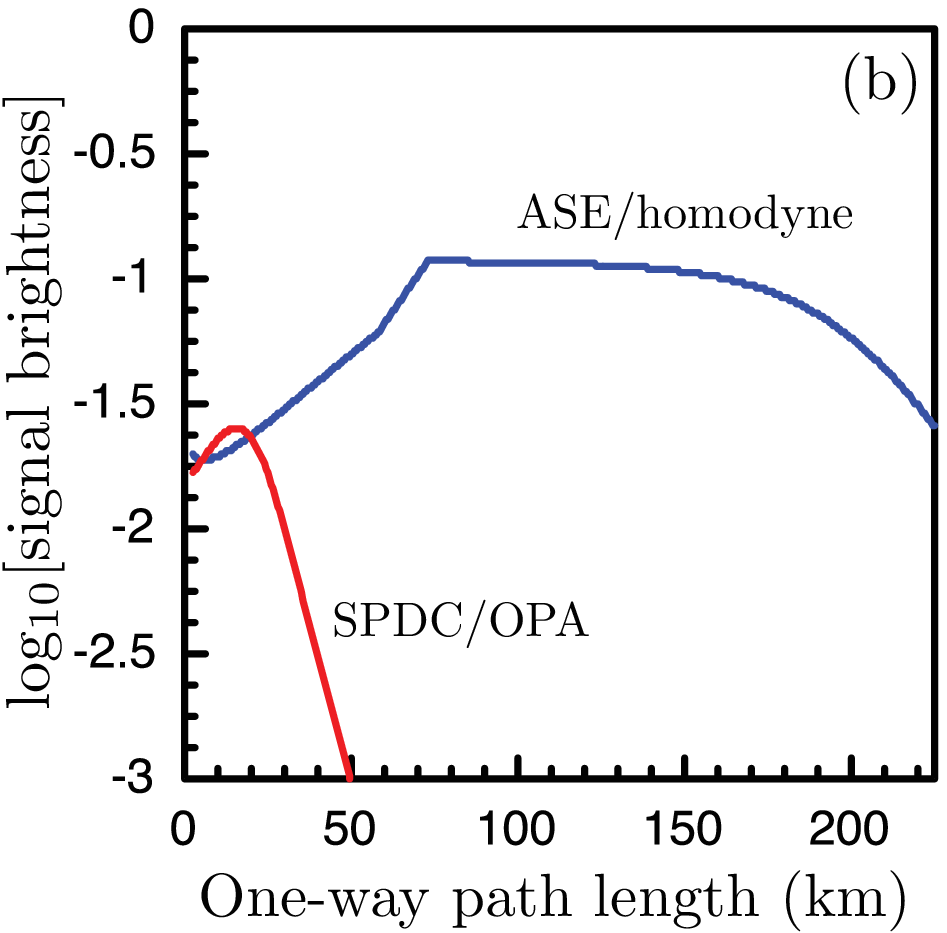}
\caption{(color online). (a) Lower bounds on Alice and Bob's secure rates when Eve mounts a collective passive-eavesdropping attack.  (b) Signal brightness, $N_S$, at which these secure rates are realized.  See text for additional details.}
\end{figure} 

There are several points worthy of note regarding the results shown in Fig.~2(a). The first is a reminder that Alice and Bob's passive-eavesdropping immunity applies to QSDC, in addition to QKD.  The second is that obviating Alice's storage loss by use of ASE light provides an enormous increase in secure-rate-versus-distance performance.  Indeed, the ASE/homodyne system is capable of a 3.5\,Gbps secure rate over 50\,km of low-loss fiber, whereas the SPDC/OPA system's secure rate is about $10^3$-times lower at that distance.  A third, more fundamental, point concerns the physical basis for this passive eavesdropping immunity.  Prior work ascribed QI's passive-eavesdropping immunity to the entanglement between the signal and idler beams produced by Alice's SPDC source \cite{Shapiro2009a,Zhang2013,Shapiro2014}, making its experimental demonstration the first realized example of entanglement's benefit surviving an entanglement-breaking channel.  Alice's switching to an ASE source to provide both her low-brightness signal and high-brightness local oscillator implies that entanglement is \em not\/\rm\ necessary to realize immunity to passive eavesdropping.  With the ASE source, security now rests on the no-cloning theorem, i.e., Eve is unable to obtain a high-brightness version of the signal light Alice sends to Bob with which to make her own high-quality homodyne measurement of Bob's message.  Finally, we point out that, over the path lengths shown, the optimum $R\le 10\,$Gbps value \em without\/\rm\ the $\Pr(e)_{\rm Alice} \le 0.1$ constraint is 10\,Gbps.  Thus the abrupt slope change in Fig.~2(a)'s ASE/homodyne curve occurs when $R$ must be reduced to satisfy that constraint.  It follows that Alice and Bob's secure-rate/distance region can be increased if both the modulation rate and error-probability constraints are relaxed, as could be done with advances in modulator technology and further development of high-efficiency error-correcting codes.

Now let us return to the issue of defeating an active attack.  Our  proposal for defeating such attacks is shown in Fig.~3.  Alice passively combines signal light from a bandwidth $W$, brightness $N_{\rm SPDC}$  SPDC source with light from highly-attenuated, bandwidth $W$ ASE source in a 1:99 ratio, resulting in a bandwidth $W$, brightness $N_S$ signal beam that she sends to Bob.  Alice uses a single-photon counter to detect photons from her SPDC source's idler beam, and retains her bright ASE light for use as a broadband homodyne LO.  Because Alice and Bob will be monitoring the power levels and power spectra at their respective terminals, Eve minimizes her exposure and maximizes her Holevo information by:  (1) replacing the 0.2\,dB/km fibers connecting Alice and Bob with her own lossless fibers; (2) injecting signal light from an SPDC source identical to Alice's into the Alice-to-Bob channel through a beam splitter, while preserving the power level and spectrum seen by Bob; and (3) withdrawing some light from the Bob-to-Alice channel through another beam splitter, while preserving the power level and spectrum seen by Alice.  Alice and Bob's security rests on: (1) their measuring the singles rates for Alice's detection of SPDC idler photons and Bob's detection of photons he has tapped from the light he received from Alice; and (2) their monitoring the coincidence rate for these measurements.  That coincidence rate is degraded, relative to the singles rates, by Eve's light injection.  In particular, Alice and Bob's singles ($S_A$ and $S_B$) and coincidence ($C_{AB}$) rates obey \cite{footnote3} $S_A = \eta N_{\rm SDPC} W$, $S_B = \eta \kappa_B\kappa_S N_S W$, and
$C_{AB} =  S_AS_BT_g + \eta^2\kappa_Bf\kappa_S\kappa_AN_{\rm SPDC}W$.
Here: $\eta$ is the detection efficiency of Alice and Bob's monitors; $\kappa_A\sim 0.9$ is the fraction of Alice's SPDC signal light that is coupled into the fiber going to Bob; $\kappa_S = 10^{-0.02 L}$ is the transmissivity of the $L$-km-long fibers connecting Alice and Bob; $\kappa_B = 0.1$ is the fraction of the light Bob receives that he routes to his monitor; $T_g = 100$\,ps is the gate time of Alice and Bob's monitors; and $f$ is the fraction of the light received by Bob that came from Alice.   

Figure~4(a) shows a lower bound on Alice and Bob's secure rate,  $\Delta I_{AB}^{\rm LB} = \beta I_{AB} - \chi_{EB}^{\rm UB}$, in the presence of the active-eavesdropping Eve \cite{footnote5} when their monitoring ensures that $f=0.99$.  Also included is a brightness plot for the light Alice sends to Bob.  These curves were obtained assuming that:  (1) the brightness of the signal Alice sends to Bob and Bob's bit rate $R \le 10\,$Gbps are chosen to maximize their secure rate subject to the constraint that $\Pr(e)_{\rm Alice} \le 0.1$ so that $\beta = 0.94$ can be assumed; (2) Alice's homodyne receiver has an undegraded local oscillator with brightness $N_{\rm LO} = 10^4$ and efficiency 0.9; and (3) all other system aspects are ideal.     We see that 2\,Gbps QKD is possible at 50\,km range. 
\begin{figure}
\includegraphics[width=3.25in]{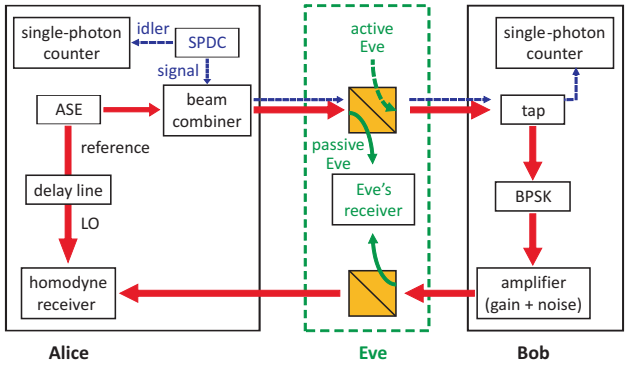}
\caption{(color online). Setup for defeating active eavesdropping by coincidence monitoring.}
\end{figure} 
\begin{figure}
\includegraphics[width=1.75in]{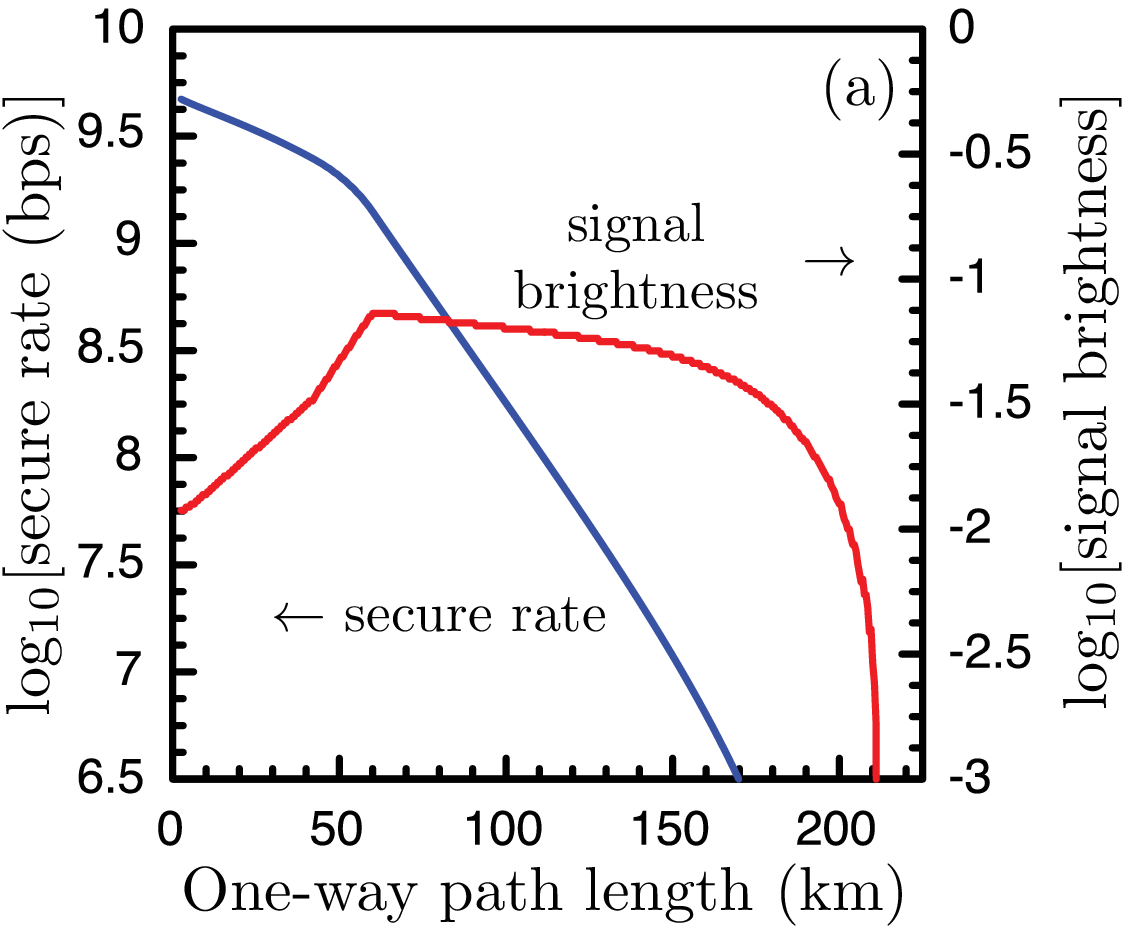}\hspace{.13in}\includegraphics[width=1.52in]{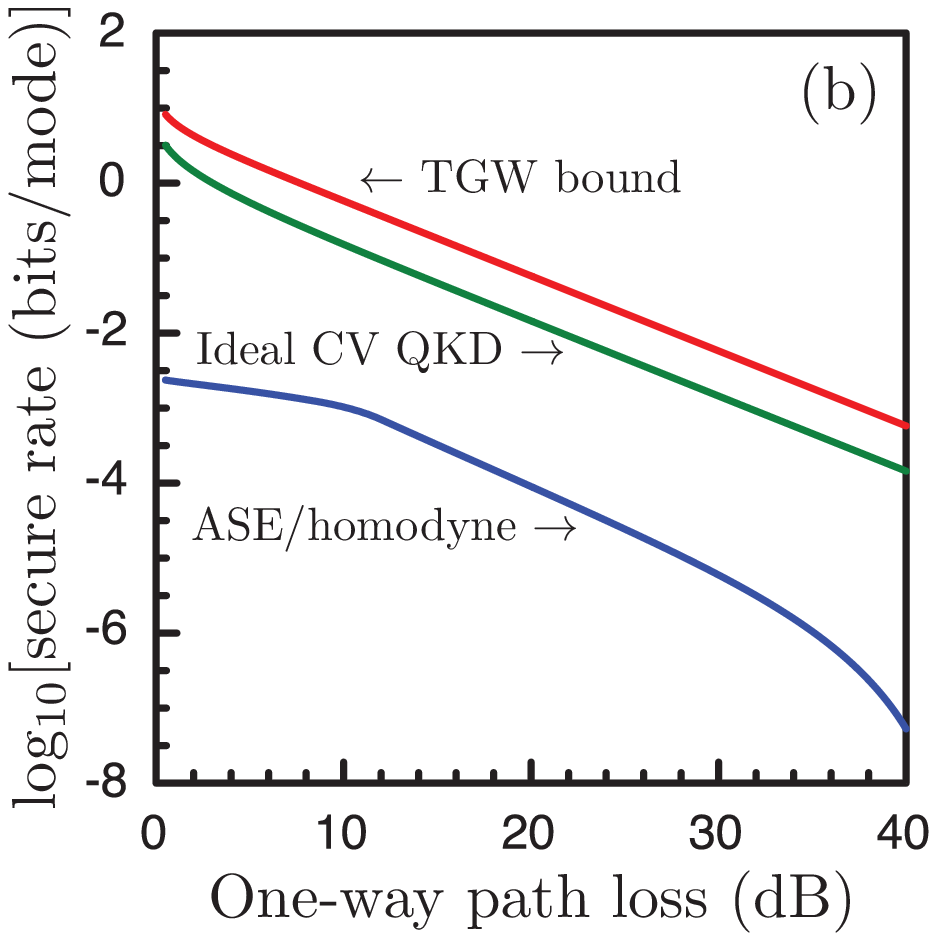}
\caption{(color online).  (a) Lower bound on Alice and Bob's secure rate when Eve mounts a collective active-eavesdropping attack, and the signal brightness, $N_S$, at which this bound is realized.  See text for additional details. (b) Secure rate in bits per mode versus one-way path loss:  comparison between the Takeoka-Guha-Wilde (TGW) bound, ideal continuous-variable (CV) QKD, and the ASE/homodyne protocol from Fig.~3 whose performance (in bits/sec) is shown in (a).}
\end{figure} 

At this juncture, there are two points worth discussing.  First, there is a major distinction to be drawn between the active and passive cases.  It will take a time $T_M$ for Alice and Bob to verify, from their measured singles and coincidence rates, that Eve's intrusion does not violate $f \ge 0.99$.  Thus QSDC operation is not possible unless Bob stores his modulated and amplified light for a $T_M$-duration interval prior to sending it to Alice, because he must ensure that the channel is safe to use for QSDC before he transmits his message.  At a 50\,km Alice-to-Bob path length,  we have found $T_M\sim 1\,$s \cite{footnote3}, making Bob's optical storage out of the question with current technology.  Absent a breakthrough in wideband, long-lived quantum memory, our approach to defeating active eavesdropping therefore restricts Alice and Bob to using their system for QKD \cite{footnote6}.  That said, their secure rate at 50\,km is orders of magnitude beyond what is possible with other QKD protocols, making it fast enough to support one-time pad file encryption.    The second point of note concerns the degree to which the active attack we have considered is the most powerful one that Eve could mount.  We can show that Eve's use of an SPDC source is her best beam-splitter active attack for a 3-mode $\chi^{\rm UB}_{EB}$ when $N_S \ll 1$ in that it nearly saturates the bound set by the entanglement-assisted Holevo capacity \cite{Zhuang2015}.  The same is true, \em without\/\rm\ the $N_S\ll 1$ restriction, for a two-mode active attack in which Eve ignores the light tapped from the Alice-to-Bob channel \cite{Zhuang2015}.  

Finally, it behooves to consider how close our protocol might come to an ultimate limit.  Takeoka \em et al\/\rm. \cite{Takeoka2014} have shown that a two-way protocol for either QSDC or QKD that is carried out over a channel whose one-way transmissivity is $\kappa_S$ has its bits/mode secure rate bounded above by $2\log_2[(1+\kappa_S)/(1-\kappa_S)]$.  It is not known, however, whether this bound is attainable.  The highest known achievable bits/mode secure rate is the $-\log_2(1-\kappa_S)$ rate of ideal continuous-variable QKD \cite{Pirandola2009,GarciaPatron2009}.  Figure~4(b) compares these rates with the bits/mode behavior of our ASE/homodyne system, whose bits/mode equals its bits/s divided by $W$, because operation is through single-mode fiber.  We see that there is a substantial gap between our protocol's bits/mode performance and the ultimate limit.  Nevertheless, our protocol's bits/s performance is orders of magnitude better than any existing QKD protocol because its modes/s value is set by the phase-matching bandwidth $W\sim$THz, whereas both continuous-variable and discrete-variable QKD protocols have modes/s values set by electrical bandwidths that are $\sim$GHz, unless wavelength-division multiplexing (WDM) is employed.  WDM can also be applied advantageously to our protocol, however, enabling it to maximize its overall secure rate---beyond what we have found for single-channel system---by operating each wavelength channel at the limit of BPSK-modulator and EDFA technology with the $\Pr(e)_{\rm Alice} = 0.1$ value needed for high-efficiency error correction.   

In conclusion, we have shown how a new two-way protocol for secure communication can overcome the QI protocol's problem of idler-storage loss limiting its rate-versus-distance capability against a passive eavesdropper, \em and\/\rm\ how our protocol's addition of coincidence-based channel monitoring can defeat active eavesdropping.  In doing so we found that entanglement is \em not\/\rm\ essential to passive eavesdropping immunity:  Eve's inability to clone a low-brightness thermal-state signal beam suffices.  Most importantly, our new protocol affords unprecedented rate-versus-distance QSDC capability against a passive eavesdropper and a similar QKD capability against an active eavesdropper.  In both cases the parameter values we have assumed for our sources, optical amplifiers, photodetectors, and error-correcting codes are achievable with available technology.

This research was supported by ONR grant number N00014-13-1-0774, AFOSR grant number FA9550-14-1-0052, and the DARPA Quiness Program through U.S. Army Research Office Grant number W31P4Q-12-1-0019.
\newpage
\widetext
\begin{center}{\Large\bf Supplemental Material}
\end{center}%

\section{Passive Eavesdropping Setups}

Figure~5 reprises the setups for our passive-eavesdropping analyses of the quantum illumination (QI) protocol from \cite{Shapiro2009a} and the new protocol. 
\begin{figure}[htb]
\includegraphics[width=3.25in]{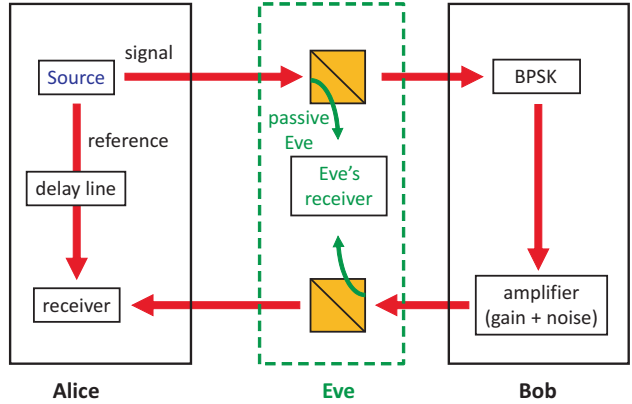}
\caption{Setups for QI-secured direct communication in the presence of passive eavesdropping.  In the protocol from \cite{Shapiro2009a}, Alice's source is a continuous-wave spontaneous parametric downconverter whose signal she sends to Bob and whose idler she retains as a reference for the optical parametric amplifier receiver from \cite{Guha2009}.  For the new protocol, Alice employs a source of amplified spontaneous emission noise to send a low-brightness signal to Bob while retaining a high-brightness copy for use as the local oscillator in a broadband balanced-homodyne receiver.}
\end{figure} 

For the QI version of this setup, Alice uses a spontaneous parametric downconverter (SPDC) source that emits $M = TW$ independent, identically-distributed, signal-idler mode pairs in each $T = 1/R$ duration bit interval with photon annihilation operators $\{\,(\hat{a}_{S_m},\hat{a}_{I_m}) : 1\le m\le M\,\}$, where $W$ is the SPDC's phase-matching bandwidth and $R$ is Bob's modulation rate.  Each such mode pair is in a zero-mean, maximally-entangled Gaussian state, i.e., a two-mode squeezed-vacuum state, that is characterized by its Wigner covariance matrix \cite{Shapiro2009a},
\begin{equation}
\Lambda_{SI} = \frac{1}{4}\left[\begin{array}{cccc}
2N_S +1 & 0&  2\sqrt{N_S(N_S+1)}& 0 \\[.05in]
0 & 2N_S + 1 & 0 & -2\sqrt{N_S(N_S+1)} \\[.05in]
2\sqrt{N_S(N_S+1)} & 0& 2N_S +1 & 0 \\[.05in]
0 & -2\sqrt{N_S(N_S+1)} & 0 & 2N_S +1 \end{array}\right],
\end{equation}
where $N_S \ll 1$.  For the new protocol, Alice uses an amplified spontaneous emission (ASE) source to obtain $M=TW$ independent, identically-distributed, signal-reference mode pairs in each bit interval with photon annihilation operators $\{\,(\hat{a}_{S_m},\hat{a}_{R_m}) : 1\le m\le M\,\}$.  The modes in each pair are in completely-correlated thermal states that are characterized by the Wigner covariance matrix,
\begin{equation}
\Lambda_{SR} =  \frac{1}{4}\left[\begin{array}{cccc}
2N_S +1 & 0&  2\sqrt{N_SN_{\rm LO}}& 0 \\[.05in]
0 & 2N_S + 1 & 0 & 2\sqrt{N_SN_{\rm LO}}\\[.05in]
2\sqrt{N_SN_{\rm LO}} & 0& 2N_{\rm LO}+1 & 0 \\[.05in]
0 & 2\sqrt{N_SN_{\rm LO}} & 0 & 2N_{\rm LO} +1 \end{array}\right],
\end{equation}
of their joint state, where $N_S \le 1 \ll N_{\rm LO}$.  

In either case, propagation of the $\{\hat{a}_{S_m}\}$ modes to Bob is through a pure-loss fiber channel with transmissivity $\kappa_S$, so that the modes he receives have annihilation operators
\begin{equation}
\hat{a}'_{S_m} = \sqrt{\kappa_S}\,\hat{a}_{S_m} + \sqrt{1-\kappa_S}\,\hat{v}_{S_m}.
\end{equation}
where the noise modes $\{\hat{v}_{S_m}\}$ are in their vacuum states.  Bob imposes his message bit, $k=0$ or $1$, on the $\{\hat{a}'_{S_m}\}$ via binary phase-shift keying (BPSK) modulation and then amplifies the modulated modes with an erbium-doped fiber amplifier (EDFA) with gain $G_B$ and output ASE $N_B$.  The modes that Bob then transmits to Alice therefore have photon annihilation operators
\begin{equation}
\hat{a}_{B_m} = (-1)^k\sqrt{G_B}\,\hat{a}'_{S_m} +  \sqrt{G_B-1}\,\hat{n}_{B_m}^\dagger,\end{equation}
where the noise modes $\{\hat{n}_{B_m}\}$ are in independent, identically-distributed, thermal states with $\langle \hat{n}_{B_m}\hat{n}^\dagger_{B_m}\rangle = N_B/(G_B-1)$.  Propagation back to Alice through a pure-loss fiber channel with transmissivity $\kappa_S$ results in a collection of return modes with photon annihilation operators given by
\begin{equation}
\hat{a}'_{B_m} = \sqrt{\kappa_S}\,\hat{a}_{B_m} +  \sqrt{1-\kappa_S}\,\hat{v}_{B_m},
\label{passive_a'Bm}
\end{equation}
where the noise modes $\{\hat{v}_{B_m}\}$ are in their vacuum states.  

For the QI setup, the idler modes are stored in a pure-loss fiber spool---with transmissivity $\kappa_I = \kappa_S^2$, yielding
\begin{equation}
\hat{a}'_{I_m} = \sqrt{\kappa_I}\,\hat{a}_{I_m} + \sqrt{1-\kappa_I}\,\hat{v}_{I_m},
\end{equation}
where the noise modes $\{\hat{v}_{I_m}\}$ are in their vacuum states.  The $\{\hat{a}'_{I_m}\}$ and the $\{\hat{a}'_{B_m}\}$ are applied as inputs to an optical parametric amplifier (OPA) whose idler-mode outputs,
\begin{equation}
\hat{a}_{{\rm out}_m} = \sqrt{G_A}\,\hat{a}'_{I_m} + \sqrt{G_A-1}\,\hat{a}_{B_m}^{'\dagger},\mbox{ for $1\le m \le M$,}
\end{equation}
are measured by quantum-efficiency $\eta$ direct detection of their photon flux over the bit interval.  Alice then decides on the value of Bob's message bit by comparing the outcome of that 
\begin{equation}
\hat{N}_{\rm OPA} = \sum_{m=1}^M \hat{a}^{'\dagger}_{{\rm out}_m}\hat{a}'_{{\rm out}_m}
\end{equation}
measurement with a threshold value chosen to equalize her false-alarm and miss probabilities \cite{Zhang2013,footnoteS1}.  If the outcome exceeds that threshold she decides Bob sent $k=0$, otherwise she decides that he sent $k=1$.  Here, we have accounted for the detector's quantum efficiency by defining
\begin{equation}
\hat{a}'_{{\rm out}_m} = \sqrt{\eta}\,\hat{a}_{{\rm out}_m} + \sqrt{1-\eta}\,\hat{v}_{{\rm out}_m},
\end{equation}
where the noise modes $\{\hat{v}_{{\rm out}_m}\}$ are in their vacuum states.

For the new ASE-based protocol, Alice detects the $\{\hat{a}'_{B_m}\}$ modes using a balanced-homodyne arrangement and decides on the value of Bob's bit by comparing the outcome of that 
\begin{equation}
\hat{N}_{\rm hom} = \sum_{m=1}^M \!\left(\hat{a}^{'\dagger}_{+m}\hat{a}'_{+m} - 
\hat{a}^{'\dagger}_{-m}\hat{a}'_{-m}\right)
\end{equation}
measurement with zero. She decides that Bob sent $k=0$ if the measurement outcome exceeds zero, and she decides $k=1$ otherwise \cite{footnoteS2}.  In this expression,
\begin{equation}
\hat{a}'_{\pm m} = \sqrt{\eta}\left(\frac{\hat{a}'_{B_m} \pm \hat{a}'_{R_m}}{\sqrt{2}}\right) + \sqrt{1-\eta}\,\hat{v}_{\pm_m},
\end{equation}
where $\eta$ is now the homodyne detector's efficiency, i.e., the product of its mode-mixing and quantum efficiencies, and the noise modes $\{\hat{v}_{\pm_m}\}$ are in their vacuum states.  

The reference modes, $\{\hat{a}_{R_m}\}$, undergo optical amplification, with gain $G_R = 1/\kappa_I$ and output ASE $N_R = G_R$, prior to being stored in a fiber spool.  That spool, which has transmissivity $\kappa_I$, has its length chosen so that its output will be delay matched to the light Alice receives from Bob, resulting in   
\begin{equation}
\hat{a}'_{R_m} = \sqrt{\kappa_I}\left(\sqrt{G_R}\hat{a}_{R_m} + \sqrt{G_R-1}\,\hat{n}^\dagger_{R_m}\right) + \sqrt{1-\kappa_I}\,\hat{v}_{R_m},
\end{equation}
with the noise modes $\{\hat{v}_{R_m}\}$ being in their vacuum states, and the noise modes $\{\hat{n}_{R_m}\}$ being in independent, identically-distributed, thermal states with $\langle \hat{n}_{R_m}\hat{n}^\dagger_{R_m}\rangle = N_R/(G_R-1)$.  For $N_{\rm LO} \gg 1$, this amplify-then-store procedure leaves the average photon number of the reference almost unchanged and it preserves nearly-complete correlation between the stored reference and the signal beam that Alice sent to Bob.  In particular, before storage we have that
\begin{equation}
\langle\hat{a}_{R_m}^\dagger\hat{a}_{R_m}\rangle = N_{\rm LO},
\end{equation}
and
\begin{equation}
\frac{|\langle \hat{a}_{S_m}^\dagger\hat{a}_{R_m}\rangle|^2}{\langle\hat{a}_{S_m}^\dagger\hat{a}_{S_m}\rangle\langle \hat{a}_{R_m}^\dagger\hat{a}_{R_m}\rangle} = 1,
\end{equation}
while after storage we find that 
\begin{equation}
\langle\hat{a}^{'\dagger}_{R_m}\hat{a}'_{R_m}\rangle = \kappa_IG_RN_{\rm LO} + \kappa_IN_R = N_{\rm LO} + 1 \approx N_{\rm LO},
\end{equation}
and
\begin{equation}
\frac{|\langle \hat{a}_{S_m}^\dagger\hat{a}'_{R_m}\rangle|^2}{\langle\hat{a}_{S_m}^\dagger\hat{a}_{S_m}\rangle\langle \hat{a}_{R_m}^{'\dagger}\hat{a}'_{R_m}\rangle} = \frac{\kappa_IG_RN_SN_{\rm LO}}{N_S(\kappa_IG_RN_{\rm LO}+\kappa_IN_R)} = \frac{N_{\rm LO}}{N_{\rm LO}+1} \approx 1
\end{equation}
when $N_{\rm LO} \gg 1$ \cite{footnoteS3}.  

\section{Active Eavesdropping Setup}

Figure~6 reprises the setup for our active eavesdropping analysis.  Now, Alice uses both SPDC and ASE sources.  For each bit interval, the SPDC source produces $M$ signal-idler mode pairs, with annihilation operators $\{\,(\hat{a}_{S_m}^{\rm SPDC},\hat{a}_{I_m}^{\rm SPDC}) : 1 \le m \le M\,\}$, that are in independent, identically-distributed, zero-mean jointly Gaussian states that are characterized by the Wigner covariance matrix
\begin{equation}
\Lambda_{SI}^{\rm SPDC} =  \frac{1}{4}\left[\begin{array}{cccc}
2N_{\rm SPDC} +1 & 0&  C_{\rm SPDC} & 0 \\[.05in]
0 & 2N_{\rm SPDC} + 1 & 0 & -C_{\rm SPDC}\\[.05in]
C_{\rm SPDC} & 0& 2N_{\rm SPDC} +1 & 0 \\[.05in]
0 & -C_{\rm SPDC} & 0 & 2N_{\rm SPDC} +1 \end{array}\right],
\end{equation}
where $N_{\rm SPDC} \ll 1$ and $C_{\rm SPDC} = 2\sqrt{N_{\rm SPDC}(N_{\rm SPDC}+1)}$.  For each bit interval, the ASE source produces $M$ signal-reference mode pairs, with annihilation operators $\{\,(\hat{a}^{\rm ASE}_{S_m},\hat{a}^{\rm ASE}_{R_m}) : 1\le m \le M\,\}$, that are in independent, identically-distribution, completely-correlated thermal states that are characterized by the Wigner covariance matrix,
\begin{equation}
\Lambda^{\rm ASE}_{SR} =  \frac{1}{4}\left[\begin{array}{cccc}
2N_{\rm ASE} +1 & 0&  2\sqrt{N_{\rm ASE}N_{\rm LO}}& 0 \\[.05in]
0 & 2N_{\rm ASE}+ 1 & 0 & 2\sqrt{N_{\rm ASE}N_{\rm LO}}\\[.05in]
2\sqrt{N_{\rm ASE}N_{\rm LO}} & 0& 2N_{\rm LO} + 1 & 0 \\[.05in]
0 & 2\sqrt{N_{\rm ASE}N_{\rm LO}} & 0 & 2N_{\rm LO} +1 \end{array}\right],
\end{equation}
where $N_{\rm ASE} = 1 \ll N_{\rm LO}$.  

\begin{figure}[htb]
\includegraphics[width=3.5in]{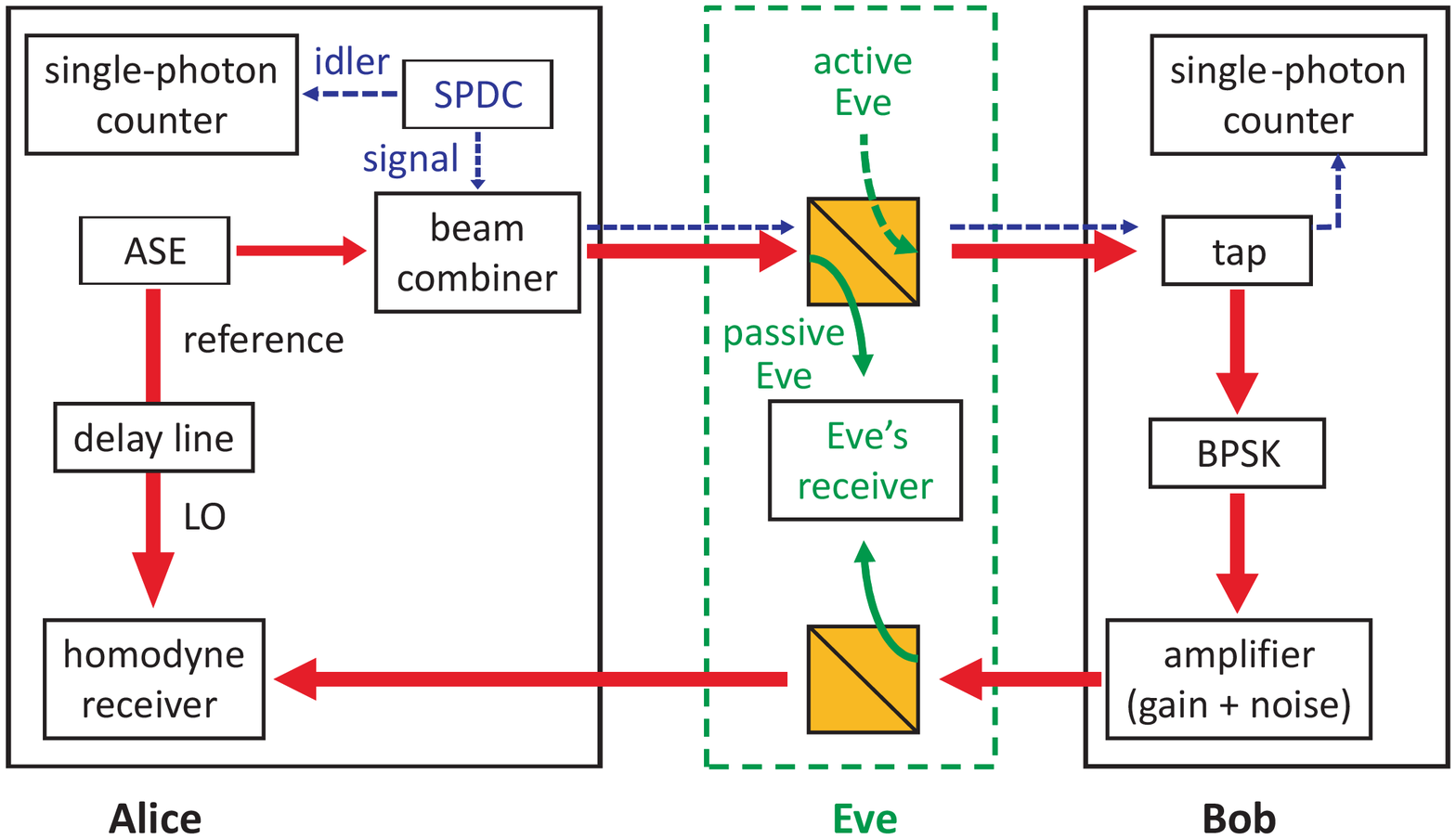}
\caption{Setup for defeating active eavesdropping by coincidence monitoring.}
\end{figure} 

Alice sends her SPDC's idler beam to her channel monitor, and combines her SPDC and ASE source's signal beams on an asymmetric beam splitter and transmits the output modes,
\begin{equation}
\hat{a}_{S_m} = \sqrt{\kappa_A}\,\hat{a}^{\rm SPDC}_{S_m} + \sqrt{1-\kappa_A}\,\hat{a}^{\rm ASE}_{S_m},
\end{equation}
to Bob.  Because she wants each of these output modes to have average photon number $N_S \le 1$, and she wants 99\% of the light going to Bob to come from her ASE source, Alice uses $\kappa_A = 1-0.99N_S$ and adjusts her downconverter's pump power to obtain $N_{\rm SPDC} = N_S/100(1-0.99N_S)$.  Note that for $N_S \le 0.1$ these choices imply $\kappa_A \ge 0.9$.  It follows that the signal modes Alice sends to Bob, her SPDC idler modes, and her ASE reference modes---denoted $\{\,(\hat{a}_{S_m},\hat{a}_{I_m},\hat{a}_{R_m}): 1\le m \le M\,\}$---are independent, identically-distributed, mode triples.  Each such mode triple is in a zero-mean Gaussian state that is completely characterized by the Wigner covariance matrix
\begin{equation}
\Lambda_{SIR} = 
\frac{1}{4}\left[\begin{array}{cccccc}
2N_S + 1& 0 & \sqrt{\kappa_A}\,C_{\rm SPDC} & 0 & C_{\rm ASE} & 0 \\[.05in]
0 & 2N_S + 1 & 0 & - \sqrt{\kappa_A}\,C_{\rm SPDC} & 0 & C_{\rm ASE} \\[.05in]
\sqrt{\kappa_A}\,C_{\rm SPDC} & 0 & 2N_{\rm SPDC}+1 & 0 & 0 & 0 \\[.05in]
 0 & - \sqrt{\kappa_A}\,C_{\rm SPDC} & 0 & 2N_{\rm SPDC}+1 & 0 & 0 \\[.05in]
C_{\rm ASE}   & 0 & 0 & 0 & 2N_{\rm LO} + 1 & 0 \\[.05in]
  0 & C_{\rm ASE}  & 0 & 0 & 0 & 2N_{\rm LO} + 1 \\[.05in]
\end{array}\right],
\end{equation} 
where $C_{\rm ASE} = 2\sqrt{(1-\kappa_A)N_{\rm ASE}N_{\rm LO}}$.

Eve replaces the lossy fibers that had connected Alice and Bob with lossless fibers into which she has inserted beam splitters with transmissivities $\kappa_{AB}$ for the Alice-to-Bob channel and $\kappa_{BA}$ for the Bob-to-Alice channel.  She has her own SPDC source (with the same phase-matching bandwidth as Alice's), whose $M$ signal-idler pairs are independent and identically distributed with each mode pair, $(\hat{a}^{\rm Eve}_{S_m},\hat{a}^{\rm Eve}_{I_m})$, being in a zero-mean, maximally-entangled Gaussian state characterized by the Wigner covariance matrix,
\begin{equation}
\Lambda^{\rm Eve}_{SI} = \frac{1}{4}\left[\begin{array}{cccc}
2N_{\rm Eve} +1 & 0&  2\sqrt{N_{\rm Eve}(N_{\rm Eve}+1)}& 0 \\[.05in]
0 & 2N_{\rm Eve} + 1 & 0 & -2\sqrt{N_{\rm Eve}(N_{\rm Eve}+1)} \\[.05in]
2\sqrt{N_{\rm Eve}(N_{\rm Eve}+1)} & 0& 2N_{\rm Eve} +1 & 0 \\[.05in]
0 & -2\sqrt{N_{\rm Eve}(N_{\rm Eve}+1)} & 0 & 2N_{\rm Eve} +1 \end{array}\right].
\end{equation}
Eve injects her SPDC's signal beam into the Alice-to-Bob channel through her $\kappa_{AB}$ beam splitter, so that the $m$th-mode signal light that Bob receives has annihilation operator
\begin{equation}
\hat{a}'_{S_m} = \sqrt{\kappa_{AB}}\,\hat{a}_{S_m} + \sqrt{1-\kappa_{AB}}\,\hat{a}^{\rm Eve}_{S_m}, 
\end{equation}
and she taps light from the Bob-to-Alice channel via her $\kappa_{BA}$ beam splitter, so that the $m$th-mode return light that Alice receives has annihilation operator
\begin{equation}
\hat{a}'_{B_m} = \sqrt{\kappa_{BA}}\,\hat{a}_{B_m} + \sqrt{1-\kappa_{BA}}\,\hat{v}_{BA_m}, 
\label{active_a'Bm}
\end{equation}
where the noise mode $\hat{v}_{BA_m}$ is in its vacuum state.  Now, because Bob uses a transmissivity-$\kappa_B$ beam splitter to divert part of the light he receives from Alice to his channel monitor, $\hat{a}_{B_m}$ in Eq.~(\ref{active_a'Bm}) is given by
\begin{equation}
\hat{a}_{B_m} = (-1)^k\sqrt{G_B}\left[\sqrt{1-\kappa_B}\,\hat{a}'_{S_m} + \sqrt{\kappa_B}\,\hat{v}'_{B_m}\right] + \sqrt{G_B-1}\,\hat{n}^\dagger_{B_m},
\end{equation}
instead of by Eq.~(\ref{passive_a'Bm}), with the noise mode $\hat{v}'_{B_m}$ being in its vacuum state.

Eve attempts to determine the full string of bits that Bob has sent by joint measurement of the light she taps from the Alice-to-Bob channel,
\begin{equation}
\hat{a}^{\rm Eve}_{AB_m} = \sqrt{1-\kappa_{AB}}\,\hat{a}_{S_m} -\sqrt{\kappa_{AB}}\,\hat{a}^{\rm Eve}_{S_m},
\end{equation}
her retained idler modes $\{\hat{a}^{\rm Eve}_{I_m}\}$, and the light she taps from the Bob-to-Alice channel,
\begin{equation}
\hat{a}^{\rm Eve}_{BA_m} = \sqrt{1-\kappa_{BA}}\,\hat{a}_{B_m} -\sqrt{\kappa_{BA}}\,\hat{v}_{BA_m},
\end{equation}
carried out over \em all\/\rm\ Bob's transmitted bits.  

Bob uses a beam splitter (whose free input-port modes are in their vacuum states) to divert a fraction $\kappa_B$ of the light he receives to his channel monitor, sending the rest to his BPSK modulator and EDFA for transmission back to Alice.  To minimize the detectability of her attack, Eve chooses $N_{\rm Eve}$, $\kappa_{AB}$, and $\kappa_{BA}$ so that:  (1)  Bob receives an average photon number $\kappa_SN_S$, just as he did in Eve's absence; (2) the fraction, $1-f$, of light entering Bob's terminal that is due to Eve's injection is 0.01; and (3) Alice receives an average photon number $\kappa_S[G_B(1-\kappa_B)\kappa_SN_S + N_B]$, just as she did in Eve's absence.  These objectives are satisfied when
$\kappa_{AB}N_S + (1-\kappa_{AB})N_{\rm Eve} = \kappa_SN_S$,
$f \equiv 1-(1-\kappa_{AB})N_{\rm Eve}/\kappa_SN_S =0.99$,
and $\kappa_{BA} = \kappa_S$.
Alice makes her decision about Bob's message bit by comparing the outcome of her homodyne measurement---as described in Sec.~I---with zero, and deciding that Bob sent $k=0$ if the measurement outcome exceeds zero and deciding $k=1$ otherwise.  

\section{Alice's Error Probabilities and Alice and Bob's Shannon Information Rates}

Because $M\ge 200$ for all the performance evaluations presented in the paper, we can use the Central Limit Theorem to justify the following Gaussian-approximation formulas for Alice's error probabilities \cite{Zhang2013} when Bob's bit value is equally likely to be 0 or 1 and Eve mounts a passive attack against an SPDC/OPA system or an ASE/homodyne system:
\begin{align}
\Pr(e)_{\rm Alice}^{\rm OPA} &= Q\!\left(\frac{\mu^{\rm OPA}_0 - \mu^{\rm OPA}_1}{\sigma^{\rm OPA}_0 + \sigma^{\rm OPA}_1}\right),\\[.05in]
\Pr(e)_{\rm Alice}^{\rm ASE} &= Q\!\left(\frac{\mu^{\rm ASE}_0 - \mu^{\rm ASE}_1}{\sigma^{\rm ASE}_0 + \sigma^{\rm ASE}_1}\right),
\end{align}
where   
\begin{equation}
Q(x) = \int_x^\infty\!{\rm d}t\,\frac{e^{-t^2/2}}{\sqrt{2\pi}}.
\end{equation}
Here, $\mu^{\rm OPA}_k$ and $\sigma^{\rm OPA}_k$ are the conditional mean and conditional standard deviation of the $\hat{N}_{\rm OPA}$ measurement given the value of Bob's message bit, $k$, when Eve makes a passive attack.  Likewise $\mu^{\rm ASE}_k$ and $\sigma^{\rm ASE}_k$ are the conditional mean and conditional standard deviation of the $\hat{N}_{\rm hom}$ measurement given the value of Bob's message bit, $k$, when Eve makes a passive attack.  A similar Gaussian-approximation formula applies when Bob's bit values are equally likely to be 0 or 1 and Eve makes an active attack:
\begin{equation}
\Pr(e)_{\rm Alice}^{\rm act} = Q\!\left(\frac{\mu^{\rm act}_0 - \mu^{\rm act}_1}{\sigma^{\rm act}_0 + \sigma^{\rm act}_1}\right),
\end{equation}
with $\mu^{\rm act}_k$ and $\sigma^{\rm act}_k$ being the conditional mean and conditional standard deviation of the $\hat{N}_{\rm hom}$ measurement given the value of Bob's message bit, $k$, when Eve makes an active attack and Alice and Bob perform channel monitoring. 

Once Alice's error probabilities are found, Alice and Bob's Shannon-information rates (in bps) follow immediately from
\begin{equation}
I_{AB} = R\{1+\Pr(e)_{\rm Alice}\log_2[\Pr(e)_{\rm Alice}] + (1-\Pr(e)_{\rm Alice})\log_2[1-\Pr(e)_{\rm Alice}]\},
\end{equation}
hence all that remains is to determine the conditional means and standard deviations needed to instantiate our error-probability formulas.  These conditional moments are easily calculated from the state characterizations and annihilation-operator transformations  in the first two sections of this Supplemental Material, so we will merely present the results.  For passive eavesdropping on the SPDC/OPA system we have
\begin{align}
\mu^{\rm OPA}_k &= M\eta\left\{G_A\kappa_IN_S + (G_A-1)[\kappa_S(G_B\kappa_SN_S + N_B) + 1] \right.\nonumber\\[.05in]
&\hspace{.1in} + \left.2(-1)^k\sqrt{G_A(G_A-1)\kappa_I\kappa_S^2G_BN_S(N_S+1)}\right\},\\[.05in]
\sigma^{\rm OPA}_k &= \sqrt{\mu^{\rm OPA}_k(\mu^{\rm OPA}_k/M +1)},
\end{align} 
while for passive eavesdropping on the ASE/homodyne system we have
\begin{align}
\mu^{\rm ASE}_k &= 2(-1)^kM\eta \kappa_S\sqrt{G_BN_SN_{\rm LO}}, \\[.05in]
\sigma^{\rm ASE}_k & = \sqrt{M\{\eta(\kappa_S^2G_BN_S + \kappa_SN_B + N_{\rm LO}) + 2\eta^2(2\kappa_S^2G_BN_S + \kappa_SN_B)N_{\rm LO}\}}, 
\end{align}
where perfect reference storage has been assumed \cite{footnoteS4}.  At this point we can obtain the asymptotic ($N_B, N_{\rm LO} \gg 1$) form of $\Pr(e)^{\rm ASE}_{\rm Alice}(e)$ that was used for illustrative purposes in the paper, albeit not in the performance-evaluation figures.  In this asymptotic regime we have that
\begin{equation}
\sigma^{\rm ASE}_k \longrightarrow \sqrt{2M\eta^2\kappa_SN_BN_{\rm LO}},
\end{equation}
whence
\begin{equation}
\Pr(e)^{\rm ASE}_{\rm Alice} \longrightarrow Q\!\left(\sqrt{2M\kappa_SG_BN_S/N_B}\right) \le 
\exp(-W\kappa_SG_BN_S/RN_B)/2,
\end{equation}
where we have used the well-known bound $Q(x) \le \exp(-x^2/2)/2$ for $x\ge 0$, and $M = W/R$.  In the paper, this asymptotic expression was quoted for ideal equipment, which presumes unity homodyne efficiency ($\eta =1$).  The derivation we have just given verifies that in this asymptotic regime $\Pr(e)^{\rm act}_{\rm Alice}$ is not sensitive to the homodyne efficiency.  Thus the $\eta = 0.9$ homodyne efficiency assumed in the paper is \em not\/\rm\ a critical value.

For active eavesdropping on the ASE/homodyne system with channel monitoring we have the
\begin{align}
\mu^{\rm act}_k &= 2(-1)^kM\eta \sqrt{\kappa_{BA}G_B(1-\kappa_B)\kappa_{AB}(1-\kappa_A)N_{\rm ASE}N_{\rm LO}}, \\[.05in]
\sigma^{\rm act}_k & = \sqrt{M\{\eta[N^{\rm act}_R + N_{\rm LO}] + 2\eta^2[N^{\rm act}_RN_{\rm LO} + \kappa_{BA}G_B(1-\kappa_B)\kappa_{AB}(1-\kappa_A)N_{\rm ASE}N_{\rm LO}]\}}, 
\end{align}
where $N^{\rm act}_R = \kappa_{BA}G_B(1-\kappa_B)\kappa_SN_S + \kappa_{BA}N_B$ and perfect reference storage has been assumed \cite{footnoteS4}.

\section{Upper Bounds on Eve's Holevo Information Rates}
For any of our configurations---passive eavesdropping on the SPDC/OPA system, passive eavesdropping on the ASE/homodyne system, and active eavesdropping on the ASE/homodyne system with channel monitoring---Eve's Holevo information rate is given by
\begin{equation}
\chi_{EB} = \left[S({\boldsymbol \rho}_E) -\sum_{k=0}^1S({\boldsymbol \rho}^{(k)}_E)/2\right]\!R,
\end{equation}
where $S(\cdot)$ denotes von Neumann entropy, ${\boldsymbol \rho}^{(k)}_E$ is Eve's conditional joint density operator for the modes she collects that are associated with a single message bit from Bob when that bit's value is $k$, and ${\boldsymbol \rho}_E = \sum_{k=0}^1{\boldsymbol \rho}^{(k)}_E/2$ is her unconditional joint density operator for those modes.  To place an upper bound on $\chi_{EB}$ we will follow the procedure used in \cite{Zhang2013,Shapiro2014} for the passive eavesdropping attack.  There, because the conditional joint density operators were Gaussian states, their von Neumann entropies were evaluated exactly by symplectic diagonalization of the conditional Wigner covariances ${\boldsymbol \Lambda}^{(k)}_E$  \cite{Pirandola2008}.  The unconditional joint density operator, however, was \em not\/\rm\ a Gaussian state, but it was zero-mean with a diagonal Wigner covariance matrix ${\boldsymbol \Lambda}_E = \sum_{k=0}^1{\boldsymbol \Lambda}^{(k)}_E/2$, meaning that all its modes were uncorrelated.  The joint state with that covariance matrix which has the highest von Neumann entropy is a collection of independent thermal states whose average photon numbers are directly related to those diagonal elements.  Moreover, Bob's $R$\,bps information rate is an upper bound on Eve's Holevo information rate, i.e., she cannot glean more information than what Bob sends.  Taken together, these results lead to the following upper bound on Eve's Holevo information rate for the passive eavesdropping attack \cite{Zhang2013,Shapiro2014}:
\begin{equation}
\chi_{EB}^{\rm UB} = \min\!\left[S_{\rm therm}({\boldsymbol \Lambda}_E) -\sum_{k=0}^1S({\boldsymbol \rho}^{(k)}_E)/2,1\right]\!R,
\label{chiUB}
\end{equation}
where $S_{\rm therm}({\boldsymbol \Lambda}_E)$ is the von Neumann entropy of a multi-mode thermal state with covariance matrix ${\boldsymbol \Lambda}_E$.  

For the passive eavesdropping configurations considered in the present paper, Eve collects all the light lost in propagation between Alice and Bob, i.e., for each of Bob's message bits she has access to $M$ mode pairs, $\{\,(\hat{a}^{\rm Eve}_{AB_m},\hat{a}^{\rm Eve}_{BA_m}) : 1\le m \le M\,\}$.  Given Bob's bit value is $k$, these mode pairs are in independent, identically-distributed, zero-mean Gaussian states, implying that
\begin{equation}
S({\boldsymbol \rho}^{(k)}_E) = MS(\rho^{(k)}_E),
\end{equation}
where $\rho^{(k)}_E$ is the common conditional joint-density operator for a mode pair.  That mode-pair conditional density operator is completely characterized by the Wigner covariance matrix
\begin{equation}
\Lambda^{(k)}_{AB} = \frac{1}{4}\!\left[\begin{array}{cccc}
2(1-\kappa_S)N_S + 1 & 0 & (-1)^kC_{AB}^{\rm pass} & 0 \\[.05in]
0 & 2(1-\kappa_S)N_S + 1 & 0 & (-1)^kC_{AB}^{\rm pass} \\[.05in]
(-1)^kC_{AB}^{\rm pass} & 0 & 2N_{BA}^{\rm pass} + 1 & 0\\[.05in]
0 & (-1)^kC_{AB}^{\rm pass} & 0 & 2N_{BA}^{\rm pass}+ 1 \end{array}\right],
\end{equation} 
where $C_{AB}^{\rm pass} = 2(1-\kappa_S)\sqrt{G_B\kappa_S}\,N_S$ and $N_{BA}^{\rm pass}  = (1-\kappa_S)(G_B\kappa_SN_S + N_B)$.  Moreover, averaging over Bob's equally-likely message-bit distribution, we have that ${\boldsymbol \rho}_E$ has a $4M\times 4M$ Wigner covariance matrix that is block diagonal with common $4\times 4$ blocks on its diagonal that are themselves diagonal and given by
\begin{equation}
\Lambda_{AB} = \sum_{k=0}^1 \Lambda_{AB}^{(k)}/2 = 
\frac{1}{4}{\rm diag}\!\left[\begin{array}{cccc}2(1-\kappa_S)N_S +1 & 2(1-\kappa_S)N_S+1 & 2N_{BA}^{\rm pass} + 1 & 2N_{BA}^{\rm pass}+1 \end{array}\right].
\end{equation} 
With these covariance matrices in hand, the procedure from \cite{Zhang2013,Shapiro2014} that was described above can be used to evaluate the $\chi_{EB}^{\rm UB}$ result from Eq.~(\ref{chiUB}).

The process to evaluate $\chi_{EB}^{\rm UB}$ for an active attack is slightly more complicated, because for each of Bob's message bits Eve now has access to $M$ mode triples, $\{\,(\hat{a}^{\rm Eve}_{AB_m},\hat{a}^{\rm Eve}_{I_m},\hat{a}^{\rm Eve}_{BA_m}) : 1 \le m \le M\,\}$.  Given Bob's bit value, $k$, these mode triples are in independent, identically-distributed, zero-mean Gaussian states that are characterized by the Wigner covariance matrix,
\begin{equation}
\Lambda^{(k)}_{AIB} = \frac{1}{4}\!\left[\begin{array}{cccccc}
2N_{AB}^{\rm act} + 1 & 0 & -C_{AI}^{\rm act}& 0 & (-1)^kC_{AB}^{\rm act} & 0 \\[.05in]
0 &  2N_{AB}^{\rm act} + 1 & 0 & C_{AI}^{\rm act} & 0 & (-1)^kC_{AB}^{\rm act} \\[.05in]
-C_{AI}^{\rm act} & 0 & 2N_{\rm Eve} + 1 & 0 & (-1)^kC_{IB}^{\rm act} & 0\\[.05in]
0 & C_{AI}^{\rm act} & 0 & 2N_{\rm Eve} + 1 & 0 & (-1)^{k+1}C_{IB}^{\rm act} \\[.05in]
(-1)^kC_{AB}^{\rm act} & 0 &(-1)^kC_{IB}^{\rm act} & 0 &2N^{\rm act}_{BA} + 1 & 0\\[.05in]
0 &(-1)^kC_{AB}^{\rm act} & 0 & (-1)^{k+1}C_{IB}^{\rm act} & 0 & 2N^{\rm act}_{BA} + 1\\[.05in] 
 \end{array}\right],
 \end{equation}
where 
\begin{align}
N_{AB}^{\rm act} &= [(1-\kappa_{AB})N_S + \kappa_{AB}N_{\rm Eve}],\\[.05in]
C_{AI}^{\rm act} &= 2\sqrt{\kappa_{AB}N_{\rm Eve}(N_{\rm Eve} + 1)},\\[.05in]
C_{AB}^{\rm act} &= 2\sqrt{(1-\kappa_{BA})G_B(1-\kappa_B)(1-\kappa_{AB})\kappa_{AB}}\,(N_S - N_{\rm Eve}),\\[.05in]
C_{IB}^{\rm act} &= 2\sqrt{(1-\kappa_{BA})G_B(1-\kappa_B)(1-\kappa_{AB})N_{\rm Eve}(N_{\rm Eve}+1)},\\[.05in]
N_{BA}^{\rm act} &= (1-\kappa_{BA})[G_B(1-\kappa_B)\kappa_SN_S + N_B].
\end{align}  
The conditional von Neumann entropies that we need are obtained by symplectic diagonalization of the $\Lambda_{AIB}^{(k)}$.  
Averaging over Bob's equally-likely message-bit distribution, we have that ${\boldsymbol \rho}_E$ has a $6M\times 6M$ Wigner covariance matrix that is block diagonal with common $6\times 6$ blocks on its diagonal given by
\begin{equation}
\Lambda_{AIB} = \sum_{k=0}^1 \Lambda_{AIB}^{(k)}/2 = \frac{1}{4}\!\left[\begin{array}{cccccc}
2N_{AB}^{\rm act} + 1 & 0 & -C_{AI}^{\rm act}& 0 & 0 & 0 \\[.05in]
0 &  2N_{AB}^{\rm act} + 1 & 0 & C_{AI}^{\rm act} & 0 & 0 \\[.05in]
-C_{AI}^{\rm act} & 0 & 2N_{\rm Eve} + 1 & 0 & 0 & 0\\[.05in]
0 & C_{AI}^{\rm act} & 0 & 2N_{\rm Eve} + 1 & 0 & 0 \\[.05in]
0 & 0 & 0 & 0 &2N^{\rm act}_{BA} + 1 & 0\\[.05in]
0 & 0 & 0 & 0 & 0 & 2N^{\rm act}_{BA} + 1\\[.05in] 
 \end{array}\right].
\end{equation}
This covariance matrix indicates that the $\hat{a}^{\rm Eve}_{AB_m}$ and $\hat{a}^{\rm Eve}_{I_m}$ modes have a phase-sensitive cross correlation, but they are both uncorrelated with the $\hat{a}^{\rm Eve}_{BA_m}$ mode.  Performing a symplectic diagonalization of the submatrix associated with the correlated modes, we can obtain the average photon numbers of three uncorrelated modes, the sum of whose thermal-state von Neumann entropies is an upper bound on $S({\boldsymbol \rho}_E)$.  In this manner we can complete evaluating the $\chi_{EB}^{\rm UB}$ result from Eq.~(\ref{chiUB}) for the active attack.  

\section{Channel Monitoring}
Alice and Bob use channel monitoring---measuring their singles rates, $S_A$ and $S_B$, and their coincidence rate, $C_{AB}$, to constrain the fraction of light entering Bob that comes from Eve's active attack.  Their monitors will be assumed to have detectors with quantum efficiencies $\eta$ and jitter-limited coincidence-gate times $T_g \sim 100$\,ps.  When the average number of photons illuminating each monitor in a gate time is much smaller than one---as will be the case for our performance evaluation---the average values of the preceding rates can be taken to be \cite{Shapiro2009b}
\begin{equation}
S_K  = \frac{\eta}{T_g}\int_0^{T_g}\!{\rm d}t\,\langle\hat{E}_K^{{\rm mon}\dagger}(t)\hat{E}^{\rm mon}_K(t)\rangle,\mbox{ for $K = A,B$},
\end{equation}
and
\begin{equation}
C_{AB} = \frac{\eta^2}{T_g}\int_0^{T_g}\!{\rm d}t\int_0^{T_g}\!{\rm d}u\,\langle \hat{E}_A^{{\rm mon}\dagger}(t)\hat{E}^{\rm mon}_A(t)\hat{E}_B^{{\rm mon}\dagger}(u)\hat{E}^{\rm mon}_B(u)\rangle,
\end{equation}
where $\hat{E}^{\rm mon}_K(t)$, for $K=A,B$. are the positive-frequency, $\sqrt{\mbox{photons/sec}}$-units field operators entering Alice and Bob's monitors.  For phase-matching bandwidths of interest, we will have $T_gW \gg 1$, so that we can replace the average photon-flux integrals in the preceding expressions with sums of photon-number operators for those fields' Fourier-series decompositions on the $[0,T_g]$ time interval.  We thus obtain
\begin{equation}
S_K = \frac{\eta}{T_g}\sum_{m=1}^{M_g} \langle \hat{a}_{K_m}^{{\rm mon}\dagger}\hat{a}^{\rm mon}_{K_m}\rangle,
\end{equation}
and
\begin{equation}
C_{AB} = \frac{\eta^2}{T_g}\sum_{m=1}^{M_g}\sum_{n=1}^{M_g}\langle \hat{a}_{A_m}^{{\rm mon}\dagger}\hat{a}^{\rm mon}_{A_m}\hat{a}_{B_n}^{{\rm mon}\dagger}\hat{a}^{\rm mon}_{B_n}\rangle,
\end{equation}
where $M_g = T_gW$.  The mode pairs $\{\,(\hat{a}^{\rm mon}_{A_m},\hat{a}^{\rm mon}_{B_m}) : 1\le m \le M_g\,\}$ are in statistically independent, identically distributed, zero-mean Gaussian states that are characterized by the Wigner covariance matrix
\begin{equation}
\Lambda_{AB}^{\rm mon} = \frac{1}{4}\left[\begin{array}{cccc}
2N_{\rm SPDC} +1 & 0&  \sqrt{\kappa_Bf\kappa_S\kappa_A}\,C_{\rm SPDC} & 0 \\[.05in]
0 & 2N_{\rm SPDC} + 1 & 0 & -\sqrt{\kappa_Bf\kappa_S\kappa_A}\,C_{\rm SPDC}\\[.05in]
\sqrt{\kappa_Bf\kappa_S\kappa_A}\,C_{\rm SPDC} & 0& 2\kappa_B\kappa_SN_S +1 & 0 \\[.05in]
0 & -\sqrt{\kappa_Bf\kappa_S\kappa_A}\,C_{\rm SPDC} & 0 & 2\kappa_B\kappa_SN_S +1 \end{array}\right],
\end{equation}
from which we immediately get
\begin{align}
S_A &= \eta N_{\rm SPDC}W,\\[.05in]
S_B &= \eta \kappa_B\kappa_SN_SW, 
\end{align}
and, after use of Gaussian-state moment factoring \cite{Shapiro1994},
\begin{equation}
C_{AB} = S_AS_BT_g + \eta^2\kappa_Bf\kappa_S\kappa_AN_{\rm SPDC}W.
\end{equation}

Of course, Alice and Bob cannot exactly measure these ensemble average rates.  Instead, they rely on time-average singles-rates and coincidence-rate measurements---$\tilde{S}_A(T_M), \tilde{S}_B(T_M),$ and $\tilde{C}_{AB}(T_M)$---made over a $T_M$-s-long measurement interval, where $T_M \gg T_g$.   These are all unbiased estimators, i.e., $\langle \tilde{S}_A(T_M) \rangle = S_A$, $\langle \tilde{S}_B(T_M) \rangle = S_B$, and $\langle \tilde{C}_{AB}(T_M) \rangle = C_{AB}$.  Furthermore, the signal-to-noise ratios (SNRs) of the singles-rate measurements, 
\begin{equation}
{\rm SNR}_K = \frac{\langle \tilde{S}_K\rangle^2}{{\rm Var}(\tilde{S}_K)} \approx S_KT_M,\mbox{ for $K=A,B$},
\end{equation}
will both be much greater than unity when $T_M \ge 0.1$\,s for the parameter values of interest.  For example, with $\eta = 0.9$, $N_{\rm SPDC} = 10^{-3}$, and $W = 2\,$THz, we have ${\rm SNR}_A \ge 82.5\,$dB when $T_M \ge 0.1$\,s.  Likewise, for $\eta = 0.9$, $N_S = 0.1$, $W = 2\,$THz, $\kappa_S = 10^{-2}$, and $\kappa_B = 0.1$, we have ${\rm SNR}_B \ge  72.5\,$dB when $T_M\ge 0.1$\,s.  Thus in assessing how much time Alice and Bob will need to ensure that Eve is constrained to $1 - f = 0.01$ fractional illumination of Bob's terminal, we will assume that their singles-rate measurements are perfect.  So, Alice and Bob can then estimate $f$ using
\begin{equation}
\tilde{f}(T_M) = \frac{(\tilde{C}_{AB}(T_M) - S_AS_BT_g)N_S}{\eta\kappa_AN_{\rm SPDC}S_B},
\end{equation}
where all of the parameters entering this estimator can be accurately calibrated.  

It is easy to see that $\tilde{f}(T_M)$ is unbiased, i.e., $\langle \tilde{f}(T_M) \rangle = f$.  To ensure that Alice and Bob can accurately decide that $f=0.99$, we shall require that $\tilde{f}(T_M)$'s standard deviation be $\delta f \le 10^{-3}$. We have that
\begin{equation}
\delta f = \frac{\sqrt{{\rm Var}[\tilde{C}_{AB}(T_M)]}\,N_S}{\eta\kappa_AN_{\rm SPDC}S_B}.
\label{deltaf}
\end{equation}
The variance term on the right satisfies
\begin{equation}
{\rm Var}[\tilde{C}_{AB}(T_M)] = 
\frac{1}{T_gT_M}{\rm Var}\!\left(\sum_{m=1}^{M_g}\sum_{n=1}^{M_g}\hat{a}^{{\rm mon}'\dagger}_{A_m}\hat{a}^{{\rm mon}'}_{A_m}\hat{a}^{{\rm mon}'\dagger}_{B_n}\hat{a}^{{\rm mon}'}_{B_n}\right),
\label{varCab}
\end{equation}
where
\begin{align}
\hat{a}^{{\rm mon}'}_{A_m} &= \sqrt{\eta}\,\hat{a}^{\rm mon}_{A_m} + \sqrt{1-\eta}\,\hat{v}^{\rm mon}_{A_m}\\[.05in]
\hat{a}^{{\rm mon}'}_{B_n} &= \sqrt{\eta}\,\hat{a}^{\rm mon}_{B_n} + \sqrt{1-\eta}\,\hat{v}^{\rm mon}_{B_n},
\end{align}
with the noise modes $\{\hat{v}^{\rm mon}_{A_m},\hat{v}^{\rm mon}_{B_n}\}$ being in their vacuum states.  
Equation~(\ref{varCab}) can be evaluated from Gaussian-state moment factoring, with the following result,
\begin{eqnarray}
{\rm Var}[\tilde{C}_{AB}(T_M)] & =&\frac{\eta^4\kappa_BN_{\rm SPDC}W}{T_M} \nonumber \\[.05in]
&\times& \left[\kappa_{AB}\kappa_A + (\kappa_SN_S + 2\kappa_{AB}\kappa_AN_{\rm SPDC})M_g + (1 + 2\kappa_B\kappa_S\kappa_A + \kappa_B\kappa_SN_S)\kappa_SM^2_gN_S\right],
\end{eqnarray}
where higher-order terms in $N_{\rm SPDC}$ and $N_S$  have been neglected because their contributions are insignificant for the parameter values of interest.  Using this result in Eq.~(\ref{deltaf}), we obtained the plot shown in Fig.~7 for the $T_M$ value needed to get $\delta f = 10^{-3}$, when Alice and Bob's monitors have quantum efficiencies $\eta = 0.9$, and all the other parameters are those given in the paper for the performance curves shown in its Fig.~4(a).
\begin{figure}[thb]
\includegraphics[width=2in]{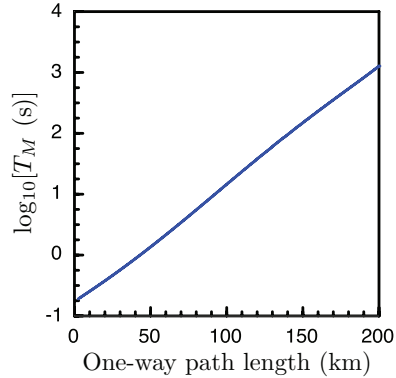}
\caption{Time duration $T_M$ needed for Alice and Bob's channel monitoring to validate $f\ge 0.99$ within a $\pm 0.001$ standard deviation.}
\end{figure}


\begin{thebibliography}{11}
\bibitem{footnote1}Here we are using an asymptotic form for this error probability---valid when $N_S \ll 1, \kappa_S \ll 1$, and $N_B \gg 1$---but in the performance examples presented later we will use more exact calculations.
\bibitem{footnote5}For passive eavesdropping, $\chi^{\rm UB}_{EB}$ is a two-mode information rate, i.e., Eve has access to light lost in Alice-to-Bob and Bob-to-Alice propagation.  For our active eavesdropping analysis, however, $\chi^{\rm UB}_{EB}$ is a three-mode information rate, viz., Eve has available the light she taps out of the Alice-to-Bob channel with her first beam splitter, her retained idler beam, and the light she taps out of the Bob-to-Alice channel with her second beam splitter.
\bibitem{footnote6}In QSDC, Bob sends an information-bearing message to Alice using an error-correcting code.  In QKD, Bob sends a random bit string to Alice after which the two use a classical channel for reconciliation, followed by the usual privacy amplification step.
\bibitem{footnoteS1}Alice's false-alarm and miss probabilities are $\Pr(e\mid k = 0)_{\rm Alice}$ and $\Pr(e\mid k = 1)_{\rm Alice}$, viz., the conditional probabilities that she makes an error given that Bob sent $k=0$ and $k=1$, respectively.
\bibitem{footnoteS3}If $\kappa_I$ is so small that amplifiers with gain $G_R = 1/\kappa_I$ are unavailable, then the fiber spool can be divided into a series connection of shorter-length spools---that together provide the required overall delay---with a loss-compensating amplifier employed at the input to each of them.  Such an arrangement can achieve the same goal of near-perfect reference storage.
\bibitem{footnoteS4}Because $\sigma^{\rm ASE}_0 = \sigma^{\rm ASE}_1$, Alice's setting her homodyne receiver's decision threshold to $(\mu^{\rm ASE}_0 + \mu^{\rm ASE}_1)/2 = 0$ minimizes $\Pr(e)_{\rm Alice}^{\rm act}$, and does so with equal false-alarm and miss probabilities.
\end{thebibliography}
\end{document}